\documentclass[prl,aps,twocolumn]{revtex4}

\usepackage{graphicx}
\usepackage{amsmath}

\newcommand{\beq}{\begin{equation}}
\newcommand{\eeq}{\end{equation}}

\begin{document}

\preprint{APS/123-QED}

\title{Diffusive redistribution of small spheres in crystallization of highly asymmetric binary hard-sphere mixtures}

\author{Wen-Sheng Xu, Zhao-Yan Sun\footnote{Correspondence author. E-mail: zysun@ciac.jl.cn}, and Li-Jia An\footnote{Correspondence author. E-mail: ljan@ciac.jl.cn}}
\affiliation{State Key Laboratory of Polymer Physics and Chemistry,
Changchun Institute of Applied Chemistry, Chinese Academy of
Sciences, Changchun 130022, People's Republic of China}



\date{\today}

\begin{abstract}
We report a molecular dynamics study of crystallization in highly asymmetric binary hard-sphere mixtures, in which the large spheres can form a crystal phase while the small ones remain disordered during the crystallization process of the large spheres. By taking advantage of assisting crystal nucleation with a patterned substrate, direct evidence is presented that there is a close link between the diffusive redistribution of the small spheres and the crystal formation of the large spheres. Although the addition of a second component with large size disparity will not alter the crystal structure formed by the large spheres, the density profile of the small spheres displays corresponding changes at different crystallization stages and closely relates to the crystal growth, suggesting possible effect of small spheres on the crystallization kinetics.
\end{abstract}


\maketitle

Crystallization of hard-sphere colloids has been the subject of intense research since assemblies of hard spheres constitute a very simple model but have potential important applications and exhibit complex behavior. Pioneering work by Alder and Wainwright in 1957~\cite{Alder} demonstrated that the monodisperse hard-sphere system crystallizes at high enough volume fraction, driven by purely entropic effects. Over the last few decades, understanding for crystallization of hard-sphere colloids has been greatly improved due to the development of new experimental techniques~\cite{vanMegen1, vanBlaaderen1, Chaikin} and novel numerical methods~\cite{Frenkel1, OMalley}. The binary hard-sphere mixtures with moderate size disparity also attract considerable interest for both practical and fundamental reasons, e.g., various types of binary crystals can be formed in such system~\cite{Dijkstra1, vanBlaaderen2, vanBlaaderen3, Schofield} and a nonstandard nucleation and growth scenario has been proposed by studying binary hard-sphere mixtures with a small to large sphere diameter ratio of $q=\sigma_{s}/\sigma_{l}=0.9$ ($\sigma_{s}$ and $\sigma_{l}$ denote small and large sphere diameters) at a total volume fraction of $0.58$~\cite{Williams}. In contrast, much less attention was received on the crystallization of highly asymmetric binary hard-sphere mixtures, in which the large spheres can form a crystal phase while the small ones remain disordered. The main reason resides in the fact that the addition of a much smaller second component will intuitively not affect the crystal structure and hence properties of the resulting material. However, the crystallization kinetics in such system is still unclear.

In this work, we focus on the crystallization of binary hard spheres with large size disparity. In such system, the presence of the small spheres can induce attractive forces between the large spheres, called the depletion effect~\cite{AO}, which is most clearly present in colloid-polymer mixtures. By taking this effect into account (through either the effective one-component model treatment or a direct study of true binary mixtures), extremely rich phase behavior, including a stable isostructural solid-solid transition, a metastable fluid-fluid transition and two types of glassy states, has been revealed in the highly asymmetric binary hard-sphere system~\cite{BHS}.  However, less is known about the role played by the smaller component in nonequilibrium processes such as crystallization process. In particular, is there a possible link between some properties of the small spheres and the crystal formation of the large spheres? In a recent paper on crystallization of colloid-polymer mixtures~\cite{Palberg}, anomalous crystallization kinetics was observed, which was postulated to be due to the diffusive redistribution of the disordered polymers. Our work can also test this postulation since the highly asymmetric binary hard-sphere mixtures can serve as a reference model for mixtures of colloids and polymers.

In principle, one can study these questions from the homogeneous crystallization of binary mixtures. However, because the nuclei are formed randomly throughout the system, it is impossible to control the orientation of the growing crystal and the shape of the forming crystallite is often irregular. This makes it difficult to character the properties of the small spheres during the crystal formation. To alleviate this, a patterned substrate can be introduced into the system as the seed of crystallization, then the direction of the growing crystal can be controlled by the substrate and the crystal grows almost layer-by-layer. Although this also introduces a degree of complexity compared to the homogeneous crystallization, we expect that the conclusion in this work can be generalized to the case of homogeneous crystallization. We find that the addition of a much smaller second component will not alter the crystal structure formed by the large spheres. However, by monitoring the density profiles of both species during the crystallization of the large spheres, we observe that the density profile for the small spheres displays corresponding changes at different crystallization stages and closely relates to the crystal growth of the large spheres.

We use a (100) patterned substrate and the lattice spacing is $a=1.576\sigma_{l}$. The patterned substrate is set in the $x$-$y$ plane and located in the middle of $z$-direction of the simulation box. Periodic boundary conditions are applied in all three directions. The spheres in the substrate are fixed, i.e., when the other spheres collide with those in the substrate, they bound off as if the spheres in the substrate have infinite mass. The previous work~\cite{Mine, Frenkel2} has shown that the forming crystal for the monodisperse hard spheres has a purely face-centered cubic (fcc) structure under the influence of this patterned substrate. For binary hard spheres with large size disparity, the structure of crystals formed by large spheres is still fcc, which will be discussed later. All binary hard-sphere liquids are quickly compressed to a large sphere volume fraction of $\phi_{l}=\pi N_{l} \sigma_{l}^{3}/6V=0.535$ (well within the coexistence region of hard spheres, $N_{l}$ stands for the large sphere number and $V$ the volume of the simulation box) by using the Lubachevsky-Stillinger (LS) algorithm~\cite{LS} with a compression rate of $0.02$. Crystallization process is then simulated using a collision-driven molecular dynamics simulation technique~\cite{Donev}. In this work, we only focus on the crystallization of binary mixtures with considerably large size ratios, and therefore the size ratio $q$ ranges from $0.1$ to $0.2$. The volume fraction of the small spheres $\phi_{s}=\pi N_{s} \sigma_{s}^{3}/6V$ with $N_{s}$ the small sphere number, covers the range from $0.001$ to $0.009$. The large and small spheres have the same mass $m$. Since supersaturation depends on the composition of binary mixtures, usage of the same compression rate will result in different supercoolings of the initial liquids. This may affect the speed of crystallization and further discussion will be given later. The temperature $T$ is set to be $1$. Length and time are given in units of $\sigma_{l}$ and $\sigma_{l}\sqrt{m/k_{B}T}$ with $k_{B}$ the Boltzmann constant. The number of large spheres is fixed at $N_{l}=N_{bulk}+N_{sub}=1000$ with $N_{sub}=100$ and $N_{bulk}=900$, where $N_{sub}$ and $N_{bulk}$ are the sphere numbers in the substrate and in the bulk. The corresponding dimensions of the simulation box are $L_{x}=L_{y}=7.88\sigma_{l}$ and $L_{z}=15.76\sigma_{l}$. Although $N_{l}=1000$ may seem too small a number to perform simulations of hard-sphere crystallization, we note that (i) the maximum value of $ N_{s }$ is $16822$ due to the small size ratio, (ii) our aim is to identify the possible role of small spheres during crystallization process, rather than to focus on the forming crystal by the large spheres, and (iii) results on the large spheres are qualitatively consistent with the previous work~\cite{Mine}, in which a much larger system was used for the same patterned substrate.

We use a spherical harmonic method~\cite{Steinhardt, Frenkel3} to identify the large spheres in a crystalline environment. It defines particles as being ordered in terms of coherence of the spherical harmonics of a particle with its neighbors. In our work, the neighbors of the large sphere are defined as those within a cutoff distance of $1.4\sigma_{l}$ from the target sphere. We use the criterions with $d_{c}=0.5$ (This criterion identifies candidate crystalline particles as having $d_{6}>d_{c}$, please see Ref.~\cite{Mine, Frenkel3} for the definition of $d_{6}$) and $N_{c}=10$ (This criterion requires the target particle to have at least $N_{c}$ neighbors that also satisfy $d_{6}>d_{c}$).

\begin{figure}[t]
 \centering
 \includegraphics[angle=0,width=0.4\textwidth]{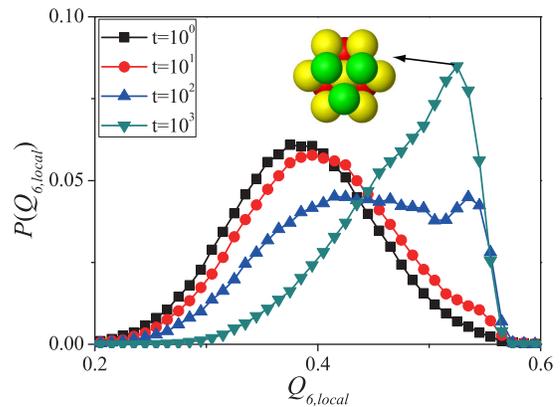}
 \caption{Sample-averaged distribution of the local bond-orientation order parameter of the large spheres at varying time for $q=0.1$ and $\phi_{s}=0.005$. The peak emerging at around $0.57$ indicates the formation of a purely fcc crystal structure. The result is averaged over $200$ independent runs.}
 \label{Fig. 1}
\end{figure}

We first present the distribution of the local bond-orientation order parameter for the large spheres at $q=0.1$ and $\phi_{s}=0.005$ in Fig. 1 (see Supplementary Video S1 for a full time series). The local bond-orientation order parameter is defined as $Q_{6,local}=(\frac{4\pi}{13}\sum^{6}_{m=-6}|\frac{1}{N_{b}^{j}}\sum^{N_{b}^{j}}_{bond=1}Y_{6m}|^{2})^{1/2}$, where $Y_{lm}$ is the spherical harmonics with $l=6$ in this case and $N_b^j$ the number of neighbors for the $j$th sphere. The values of $Q_{6, local}$ for perfect structures of fcc, hexagonal closed packed (hcp), and the body-centered cubic (bcc) crystals are $0.575$, $0.485$, $0.511$, respectively. It is clearly seen in Fig. 1 that a peak at around 0.57, which indicates the formation of the fcc structure, emerges and its height increases as crystallization proceeds. No apparent peaks appear at other positions, thus the forming crystal structure under the influence of (100) patterned substrate is purely fcc, which is consistent with the previous work~\cite{Mine, Frenkel2}. This indicates that the crystal structure formed by the large spheres is not altered by the addition of very small spheres.

\begin{figure}[t]
 \centering
 \includegraphics[angle=0,width=0.45\textwidth]{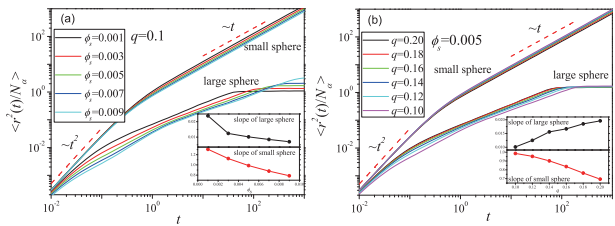}
 \caption{Sample-averaged mean squared displacements (MSD) of both species at (a) varying $\phi_{s}$ with $q=0.1$ and (b) varying $q$ with $\phi_{s}=0.005$. $N_{\alpha}$ denotes small sphere number or large sphere number in the bulk. The red dash lines with slopes $1$ and $2$ have been included to stress the fact that the small spheres undergo typical diffusion of simple liquids. The results are averaged over $50$ independent runs. Insets: extracted slopes of MSD for both species.}
 \label{Fig. 2}
\end{figure}

\begin{figure}[b]
 \centering
 \includegraphics[angle=0,width=0.45\textwidth]{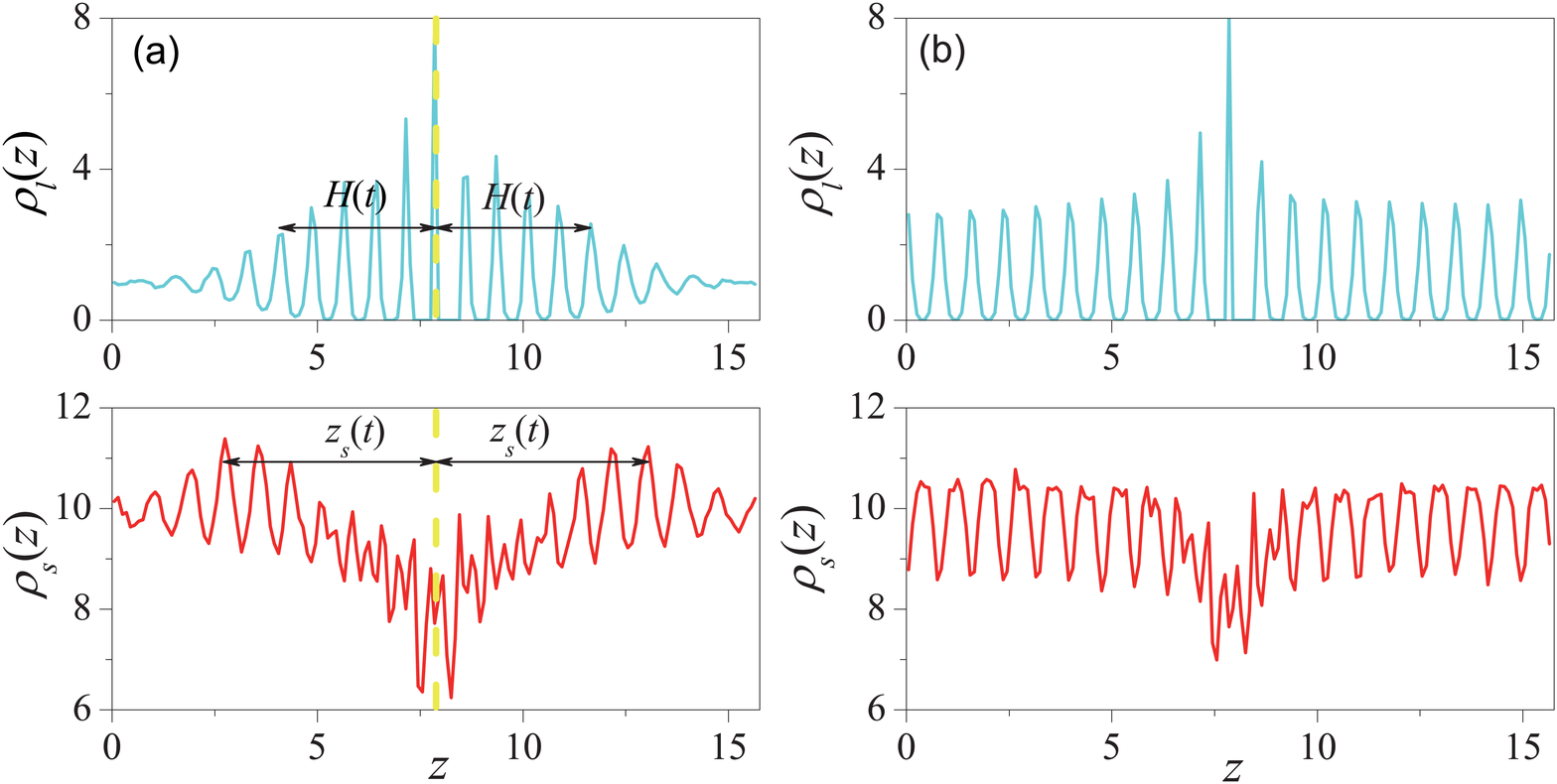}
 \caption{Sample-averaged density profiles of the large (upper panel) and small (lower panel) spheres for $q=0.1$ and $\phi_{s}=0.005$ at (a) $t=100$ and (b) $t=1000$. The yellow dash lines indicate the substrate position. $H(t)$ denotes the thickness of the crystalline film (see text for its definition) and $z_{s}(t)$ the position relative to the substrate where the maximum local density for small spheres exhibits. The density profiles are averaged over $200$ independent runs.}
 \label{Fig. 3}
\end{figure}

We show the diffusion behavior of both species in Fig. 2 by calculating the mean squared displacements (MSD). We observe that the ballistic motion ($<\Delta r^{2}(t)> \sim t^{2}$) at short times crosses into the diffusive behavior ($<\Delta r^{2}(t)> \sim t$) at long times for the small spheres. For the large spheres, the diffusive region cannot be reached since the diffusion ceases when the crystallization completes (the crystal fraction reaches $1$). At that time, a typical large sphere has diffused only about $1$ large sphere diameter. This is about $10$ times smaller than that in the homogeneous crystallization~\cite{Pusey}, indicating the high efficiency of the substrate in directing crystal growth. To better character the diffusion behavior of both species at different $q$ and $\phi_{s}$ , we also present the extracted long-time slopes of MSD in the insets. For the large spheres, the slopes are obtained by extracting long-time data before the cessation of diffusion. At fixed $q$, the slopes for both species decrease as $\phi_{s}$ increases (Fig. 2(a)). Surprisingly, the slope for large spheres increases as $q$ increases at fixed $\phi_{s}$ (Fig. 2(b)), which indicates that the speed of crystallization is fast for large size ratio at fixed volume fraction of small spheres. This is also confirmed by the growth speed of crystal fraction (data not shown). We speculate that this is because of a change in the driving force since the supersaturation of the initial supercooled binary liquids is different for various mixtures in our study. As the small spheres undergoes diffusion of typical simple liquids, one may think that the density profile of small spheres during crystallization process is fully random. However, we will show below that the distribution of the small spheres has a close connection with the crystal formation of the large spheres.

\begin{figure}[t]
 \centering
 \includegraphics[angle=0,width=0.4\textwidth]{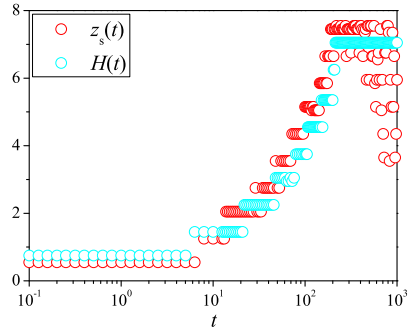}
 \caption{Connection between the position where the maximum local density for the small spheres displays and the thickness of the crystal film formed by the large spheres at $q=0.1$ and $\phi_{s}=0.005$.}
 \label{Fig. 4}
\end{figure}

In Fig. 3, we present the density profiles of both species at $t=100$ (Fig. 3(a)) and $t=1000$ (Fig. 3(b)) (see Supplementary Video S1 for a full time series). The density profiles, defined as the average number of spheres per unit volume along the $z$-axis, are accumulated in bins of width $\delta z=0.1\sigma_{l}$. Obviously, there exists a correlation between the density profiles for both species. For instance, at the stage of crystal growth ($t=100$), a maximum local density for the small spheres is found at the position where the thickness of the crystal formed by the large spheres can be roughly identified (The detailed discussion will be shown later). While, the small spheres will distribute uniformly in the voids of the resulting crystal after crystallization completes ($t=1000$).

The more clear support for connections between the small sphere distribution and the large sphere crystal formation is provided by calculating the maximum local density of the small spheres $\rho_{s}^{max}(t)$ and the thickness of the crystalline film $H(t)$ for the large spheres at time $t$. $\rho_{s}^{max}(t)$ is determined by searching for the maximum local density of the small spheres, which can be readily identified before the complete of crystallization, and the corresponding position relative to the substrate is denoted as $z_{s}(t)$. We determine $H(t)$ as the last peak position relative to the substrate in the large sphere density profile, where the peaks are defined as local densities larger than $\bar{\rho_{l}}+\rho_{t}$, where $\bar{\rho_{l}}$ indicates the average density of the large spheres along the $z$-axis and $\rho_{t}$ is a threshold. We use $\rho_{t}=1.0$ and the qualitative result is not dependent on the choice of $\rho_{t}$. The resulting $H(t)$ and $z_{s}(t)$ are shown in Fig. 4. As can be seen, they have similar behavior with $t$, i.e., the growth of the crystalline film is accompanied by a shift of $z_{s}(t)$ to large distances relative to the substrate before crystallization completes. $H(t)$ saturates at $7.05\sigma_{l}$ and $z_{s}(t)$ varies randomly when crystallization completes. Thus, the results unambiguously demonstrate a coupling between the distribution of the small spheres and the crystal growth of the large spheres during the crystallization process. In Fig. 5, we also show the maximum local density for the small spheres $\rho_{s}^{max}$ and its connection to the growth of the large sphere crystal fraction during crystallization process. We observe that (i) $\rho_{s}^{max}$ has a large value in the initial supercooled liquid and the corresponding position is near the substrate (see Fig. 4), which is due to the presence of the substrate during the compression; (ii) when entering into the stage of crystal growth ($t>1$), $\rho_{s}^{max}$ at first drops from $\sim12.2$ to $\sim11.4$ and then remains nearly constant at $\sim11.4$ for $t<100$. The latter observation suggests that there is first an expulsion of the small spheres followed by a diffusion back-in during the crystallization of the large spheres. In other words, the small spheres diffusively redistribute during the crystallization process; (iii) on the approach to the complete of crystallization, $\rho_{s}^{max}$ dramatically increases to $\sim12.0$ followed by a fast decrease, indicating the non-balance between the expulsion and back-in of the small spheres at this stage; (iv) after the complete of crystallization, the small spheres uniformly distribute in the voids of the crystal formed by the large spheres, thus $\rho_{s}^{max}$ remains unchanged at about $10.5$. The result again clearly shows that the distribution of the small spheres exhibit corresponding changes at different crystallization stages. And the observation for the small spheres may also help to better understand and distinguish the different crystallization stages.

\begin{figure}[t]
 \centering
 \includegraphics[angle=0,width=0.4\textwidth]{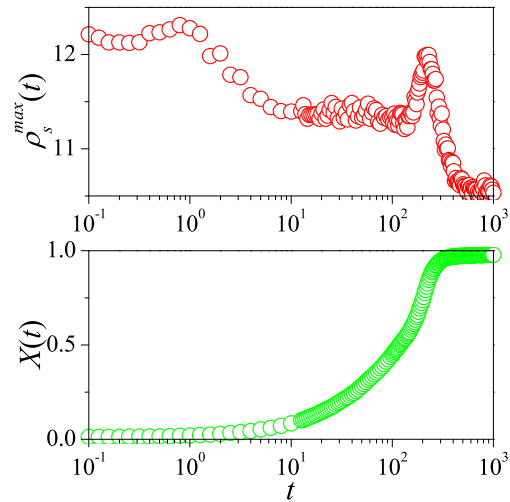}
 \caption{Connection between the maximum local density for the small spheres (upper panel) and the growth speed of crystal fraction for the large spheres (lower panel) at $q=0.1$ and $\phi_{s}=0.005$.}
 \label{Fig. 5}
\end{figure}

In summary, by taking advantage of assisting crystal nucleation with a patterned substrate, we have demonstrated very directly a close link between the density profile of the small spheres and the crystallization of the large spheres in the highly asymmetric binary hard-sphere mixtures. Our finding may potentially explain the experimentally observed anomalous crystallization kinetics in colloid-polymer mixtures. However, since our work only gives evidence that the small spheres do play some role in the crystallization process of the large spheres and the crystallization kinetics (such as the growth rate of the crystallite size) cannot be probed due to the small size in our simulations, we hope that our finding will trigger relevant experiments on highly asymmetric binary hard spheres, in which crystallization kinetics can be measured.

\begin{acknowledgments}
We thank Professor D. Frenkel for helpful discussions. This work is supported by the National Natural Science Foundation of China (21074137, 50873098, 50930001, 20734003) programs and the fund for Creative Research Groups(50921062).
\end{acknowledgments}


\end{document}